\documentclass[12pt]{article}
\pdfoutput=1
\usepackage{amsmath,amssymb,color,epsfig,draft}
\usepackage{graphicx}
\usepackage{setspace}
\usepackage{braket}
\usepackage{physics}
\usepackage{slashed}
\usepackage{simplewick}
\usepackage{cancel}
\usepackage{extarrows}
\usepackage[offset=1.25em]{simpler-wick}
\usepackage{tcolorbox}

\usepackage{tikz-cd}

\usepackage{youngtab}
\Yboxdim{4pt}
\makeatletter
\@addtoreset{equation}{section}
\makeatother

\newcommand{\bpm}{\begin{pmatrix}}
\newcommand{\epm}{\end{pmatrix}}

\def\0{{\sst{(0)}}}
\def\1{{\sst{(1)}}}
\def\2{{\sst{(2)}}}
\def\3{{\sst{(3)}}}
\def\4{{\sst{(4)}}}
\def\5{{\sst{(5)}}}
\def\6{{\sst{(6)}}}
\def\7{{\sst{(7)}}}
\def\8{{\sst{(8)}}}
\def\sst#1{{\scriptscriptstyle #1}}

\def\cH{{{\cal H}}}

\def\ie{\begin{equation}\begin{aligned}}
\def\fe{\end{aligned}\end{equation}}

\newtheorem{conj}{Conjecture}

\begin{document}

\renewcommand\arraystretch{1.5}

\begin{titlepage}

\begin{center}

\title{Fortuity and R-charge concentration in the D1--D5 CFT}

\author{Chi-Ming Chang$^{a,b}$, and Haoyu Zhang$^{c}$}

\address{${}^a$Yau Mathematical Sciences Center (YMSC), Tsinghua University, Beijing 100084, China}

\address{${}^b$Beijing Institute of Mathematical Sciences and Applications (BIMSA) \\ Beijing 101408, China}

\address{${}^c$George P. \&  Cynthia Woods Mitchell Institute for Fundamental Physics and Astronomy,
Texas A\&M University, College Station, TX 77843, USA}

\email{cmchang@tsinghua.edu.cn, zhanghaoyu@tamu.edu }

\end{center}

\vfill

\begin{abstract}

We investigate finite-$N$ BPS cohomology in the D1--D5 CFT, focusing on the sector of fortuitous classes. Analyzing the supercharge cochain complexes in the $N=2$ and $N=3$ theories, we construct several explicit fortuitous classes. We study the decomposition of these cohomology classes into ${\rm SU}(2)_a\times {\rm SU}(2)_b$ representations and conjecture that, at fixed holomorphic weight, those transforming in the largest representation are necessarily fortuitous. Our results also provide strong evidence that the $R$-charge concentration phenomenon extends to the D1--D5 CFT.

\end{abstract}

\vfill

\end{titlepage}

\section{Introduction}
Recent analyses of fortuitous phenomena in ${\cal N}=4$ SYM, the SYK model, and the ABJ vector model have shed new light on the structure of black hole microstates in AdS/CFT~\cite{Chang:2022mjp,Choi:2023znd,Budzik:2023vtr,Chang:2023zqk,Choi:2023vdm,Chang:2024zqi,Chang:2024lxt,deMelloKoch:2024pcs,Gadde:2025yoa,Chang:2025mqp,Choi:2025bhi,Kim:2025vup}. While monotone states persist to arbitrarily large $N$ and become non-interacting supergraviton states holographically, fortuitous states cease to remain BPS in the large-$N$ limit. Another well-studied holographic laboratory is the D1--D5 CFT, where the celebrated Strominger–Vafa counting of black hole microstates was first carried out~\cite{Strominger:1996sh}. The BPS spectrum of this system has been extensively investigated, both via protected indices in the boundary CFT and via supersymmetric gravitational path integrals in the bulk description~\cite{deBoer:1998kjm,deBoer:1998us,Maldacena:1999bp, Dijkgraaf:2000fq,Bena:2011zw,Benjamin:2016pil,Benjamin:2017rnd,Heydeman:2020hhw,Hughes:2025oxu,Lee:2025veh}.

To gain more refined information about black hole microstates, one approach is to determine explicit wavefunctions of BPS states in the boundary CFT by studying the lifting matrix~\cite{Gava:2002xb,Guo:2019ady,Guo:2020gxm,Benjamin:2021zkn,Gaberdiel:2023lco,Gaberdiel:2024nge,Fabri:2025rok,Gaberdiel:2025smz} in conformal perturbation theory under the exactly marginal deformation of $\mathrm{Sym}^N(T^4)$. Using the one-to-one correspondence between BPS states and supercharge cohomology classes, the second-order lifting problem reduces to computing the cohomology of the supercharge deformed at first order; we will refer to this supercharge cohomology as $Q$-cohomology.

In our earlier work~\cite{Chang:2025rqy}, we constructed several explicit $Q$-cohomology classes in the $N=2$ theory and extended the recent classification of classes~\cite{Chang:2024zqi} to the D1--D5 CFT.\footnote{See also the discussion of fortuity using the index and symmetry approach  in~\cite{Hughes:2025car}.} Building on this framework, we now present several explicit fortuitous classes in the $\tfrac{1}{4}$-BPS sector at $N=2$. Beyond the contracted large ${\cal N}=(4,4)$ quantum numbers $(h,j,\tilde h=\tilde \jmath)$, we show that the supercharge cochain complexes can be further refined by the right-moving $\mathrm{SU}(2)_{\rm outer}$ spin $\tilde\jmath_{\rm outer}$. In addition, the cohomology classes carry charges $(j_a,j_b)$ under the $\mathrm{SU}(2)_a \times \mathrm{SU}(2)_b$ symmetry~\cite{Gaberdiel:2025smz}. We decompose the cohomology classes into $\mathrm{SU}(2)_a \times \mathrm{SU}(2)_b$ representations, and, guided by our examples, conjecture that the cohomology classes transforming in the largest representations must be fortuitous.

Another distinctive feature of fortuitous classes is their R-charge concentration. Motivated by studies of Schwarzian theories, super-JT gravity~\cite{Stanford:2017thb,Heydeman:2020hhw,Turiaci:2023jfa}, and fortuity in supersymmetric SYK models, it was conjectured in \cite{Chang:2024lxt} that fortuitous classes are sharply concentrated in a single space of the $Q$-cohomology complex. Through an explicit cohomology analysis in the $N=2$ and $N=3$ theories, we provide strong evidence that this conjecture extends to the D1--D5 CFT.

\section{Fortuity at $N=2$}
We study the spectrum of $\tfrac{1}{4}$-BPS states in the D1--D5 system by analyzing the lifting problem in the conformal perturbation theory about the orbifold point $\mathrm{Sym}^N(T^4)$. 
At the orbifold point, the Hilbert space is organized by the conjugacy classes of the symmetric group $S_N$, which are labeled by cycle shapes $p$
\be
p = (1^{n_1},\, 2^{n_2},\, \dots,\, k^{n_k}, \dots),
\ee
where $n_k$ denotes the number of cycles of length $k$ and satisfies $\sum_k k\,n_k = N$.

There are two key properties of the first order deformed supercharge action:
\begin{itemize}
    \item It maps between states with different cycle structures by either splitting a single cycle into two or joining two cycles into one.
    \item It commutes with the right-moving diagonal Clifford algebra, which naturally allows the refinement of the orbifold $\frac{1}{4}$-BPS states into distinct quartets.
\end{itemize}

The classification of $Q$-cohomology classes into monotone and fortuitous sectors follows the framework of \cite{Chang:2025rqy}. The key ingredients are the supercharge action $Q_N$ at fixed $N$ and the actions of $Q_{N'}$ for all $N'>N$ on the enlarged Hilbert spaces. The Hilbert spaces at $N$ and $N'$ are related by the stringy exclusion map (SEP) $\pi_{N, N'}$, which projects orbifold BPS states at $N'$ onto those at  $N$. By analyzing how lifted states behave under $Q_{N'}$, one can distinguish monotone classes (which always admits a lift to a $Q_{N'}$ closed state at higher $N'$) from fortuitous classes.

\subsubsection*{Longest cochain complex and its lift}
In general, at a given $N$, there is more than one supercharge cochain complexes in the D1--D5 system. We chose to focus on the longest supercharge cochain complex and its lift, as it contains the maximally twisted sector and is expected to capture the largest amount of black hole entropy, making it ideal for illustrating the connection between fortuity and black hole microstates. The $N=3$ complex and its projection under $\pi_{2,3}$ to the $N=2$ complex can be summarized as
\ie
\begin{tikzcd}
 0 \arrow[r, "Q_{3}"] 
   & V_{(1^3),0}^{0} \arrow[r, "Q_{3}"] \arrow[d, "\pi_{2,3}"]  
   & V_{(2,1),0}^{1}\arrow[r, "Q_{3}"] \arrow[d, "\pi_{2,3}"]  
   & V_{(1^3),0}^{2} \oplus
 V_{(3), 0 }^{2} \arrow[r, "Q_{3}"] \arrow[d, "\pi_{2,3}"] 
    & V_{(2,1),0}^{3}\arrow[r, "Q_{3}"] \arrow[d, "\pi_{2,3}"] & V_{(1^3),0}^{4}\arrow[d, "\pi_{2,3}"]  \arrow[r, "Q_{3}"]  \arrow[d, "\pi_{2,3}"] & 0 \\
 0 \arrow[r, "Q_2"] 
   & V_{(1^2),0}^{0} \arrow[r, "Q_2"]  
   & V_{(2),0}^{1} \arrow[r, "Q_2"]  
   & V_{(1^2),0}^{2}  \arrow[r, "Q_2"]   & 0 & 0 & 
\end{tikzcd}\label{eqn:longestcochain}
\fe
where the space $V^{2\tilde{\jmath}}_{p,\,2\tilde{\jmath}_{\text{outer}}}$ is defined in Appendix~\ref{appendix: cochain_complex_def}.

\subsubsection*{Definitions of monotone and fortuitous spaces}
In \cite{Chang:2025rqy}, the $Q$-cohomology classes are classified as monotone and fortuitous. The spaces of monotone and fortuitous classes are
\ie\label{eqn:def}
\cH^{\rm mon}_N&:=(U^{\rm closed}_N+{\rm Im}(Q_N))/{\rm Im}(Q_N)\,.
\\
\cH^{\rm for}_N&:=H^*_Q(V_N)/\cH^{\rm mon}_N\cong {\rm Ker}(Q_N)\big/\big(U^{\rm closed}_N+{\rm Im}(Q_N)\big)\,.
\fe
where the space $U^{\rm closed}_N$ is defined by
\ie
U^{\rm closed}_N := {\rm Ker}(Q_N)\cap\,\bigcap_{N'>N}\,{\rm Im}\left(\pi_{N,N'}\big|_{{\rm Ker}(Q_{N'})}\right)\,.
\fe
However, directly working with these definitions requires the detailed knowledge of $Q_{N'}$-action for all $N'>N$. To make progress, a finite-cutoff proxy criterion was provided that allows determining
whether a given $Q_N$-cohomology class is approximately monotone by only studying the $Q_{N'}$-action at a fixed $N'>N$. The spaces of proxy monotone classes are defined by 
\ie
\cH_N^{\rm mon_{N'}}&:=(U^{\rm closed_{N'}}_N+{\rm Im}(Q_N))/{\rm Im}(Q_N)\,,
\\
U^{\rm closed_{N'}}_N &:= {\rm Ker}(Q_N)\cap{\rm Im}\left(\pi_{N,N'}\big|_{{\rm Ker}(Q_{N'})}\right)\,.
\fe
Note that the monotone space $\cH_N^{\rm mon}$ is a subspace of the proxy monotone space $\cH_N^{\rm mon_{N'}}$; hence, any cohomology class in the quotient $H^*_Q(V_N)/\cH_N^{\rm mon_{N'}}$ is fortuitous.

\subsubsection*{$N=2$ monotone classes}
Let us focus on the proxy monotone space $\cH_{2}^{\rm mon_{3}}$. By explicit computations, we find that all cohomology classes in $V^0_{(1^2),0}$ and $V^2_{(1^2),0}$ are proxy monotone up to $h=4$. Their degeneracies are captured by the short-character expansion (\ref{eqn:N=2_short_char_j=0}). This is consistent with the results of~\cite{Guo:2020gxm}, where it was shown that only the half-BPS states and their contracted $\mathcal{N}=(4,4)$ descendants remain unlifted. In the sector $V^1_{(2),0}$, we find that the proxy monotone degeneracies are again captured by the short-character expansion (\ref{eqn:N=2_short_char_j=1}). Therefore, the monotone up to the order we computed are contracted $\mathcal{N}=(4,4)$ descendants of the half BPS primaries. 

Each proxy monotone class we have identified is in fact a genuine monotone class in the sense of \eqref{eqn:def}. Since these proxy monotone classes arise as contracted $\mathcal{N}=(4,4)$ descendants of $\tfrac{1}{2}$-BPS states, they admit lifts to arbitrarily large $N'>2$ that are $Q_{N'}$-closed, obtained simply by lifting the contracted $\mathcal{N}=(4,4)$ algebra from $N=2$ to $N'$. Hence, we find
\ie
\cH_2^{\rm mon}= \cH_2^{\rm mon_{3}}\quad\text{up to $h=4$}\,.
\fe

\subsubsection*{$N=2$ fortuitous classes}

Quotienting out these monotone classes as in \eqref{eqn:def}, we find 26 fortuitous class at $h=2,j=0$, 608 at $h=3,j=0$, and 5922 at $h=4,j=0$ in $V_{(2),0}^1$. Focusing on the first case, we find that 20 of them transform under the $({\bf 5}, {\bf 4})$ representation of the ${\rm SU}(2)_a\times {\rm SU}(2)_b$ symmetry, in which the highest weight class has the representative:
\bea\label{eq:20hw}
\psi^-_0\alpha^1_{-\frac{1}{2}}\alpha^1_{-\frac{1}{2}}\alpha^1_{-\frac{1}{2}}\ket{2_{--,--}}
\eea
The remaining 6 fortuitous classes transform under $({\bf 2},{\bf 3})$, in which the highest weight class has the representative:
\be\label{eq:6hw}
\big[\psi^-_{-1}\alpha^1_{-\frac{1}{2}}+\frac{1}{\sqrt{3}}\psi^-_0\alpha^1_{-\frac{3}{2}}+\sqrt{2}\psi^-_0\alpha^1_{-\frac{1}{2}}\alpha^1_{-\frac{1}{2}}\bar\alpha^2_{-\frac{1}{2}}-\psi^-_0\alpha^1_{-\frac{1}{2}}\bar\alpha^1_{-\frac{1}{2}}\alpha^2_{-\frac{1}{2}}-\psi^-_{0}\psi^+_{-1}\bar\psi^-_0\alpha^1_{-\frac{1}{2}}\big] \ket{2_{--,--}}
\ee
Here, we choose the representatives to be also $ Q^\dagger$-closed; hence, \eqref{eq:20hw} and \eqref{eq:6hw} are both BPS under the second-order deformation. The ${\rm SU}(2)_a\times {\rm SU}(2)_b$ representation decomposition of the Q cohomology classes at $h=3,j=0$ is given in Appendix~\ref{app:RepDecompofh=3j=0}.

\subsubsection*{Fortuity and ${\rm SU}(2)_a\times {\rm SU}(2)_b$ symmetry}

There is an interesting interplay between the ${\rm SU}(2)_a\times {\rm SU}(2)_b$ symmetry and fortuity. Under ${\rm SU}(2)_a$, the pairs of modes $(\alpha_n^1,\bar\alpha_n^1)$, $(\alpha_n^2,\bar\alpha_n^2)$, $(\psi_r^-,\bar\psi_r^-)$, and $(\psi_r^+,\bar\psi_r^+)$ each furnish a fundamental representation. Under ${\rm SU}(2)_b$, the pairs bosonic modes $(\alpha_n^1,\alpha_n^2)$ and $(\bar\alpha_n^1,\bar\alpha_n^2)$ transform in the fundamental representation, while the fermionic modes are singlets.

With $h$ held fixed, states built from a larger number of left-moving modes transform in larger representations of the ${\rm SU}(2)_a\times {\rm SU}(2)_b$ symmetry.\footnote{Here and in the following, “larger” refers to representations of larger dimension.} Consequently, the largest ${\rm SU}(2)_a\times {\rm SU}(2)_b$ representation can only appear in the maximally twisted sector, since modes on longer cycles can carry smaller values of $h$. Moreover, within a sector of fixed cycle shape, any state in the largest representation must be constructed solely from modes living on the longest cycle in that cycle shape. These representation-theoretic constraints strongly restrict the action of the first-order deformed supercharge $Q$, which itself transforms in relatively small representations—namely, as a singlet of ${\rm SU}(2)_a$ and in a fundamental of ${\rm SU}(2)_b$, as discussed in~\cite{Gaberdiel:2025smz}.\footnote{More precisely, the first-order deformed supercharges $(Q,Q')$ form a fundamental representation of ${\rm SU}(2)_b$, where $Q'=\widetilde G'_{-1/2}$.} This leads to the following conjecture:
\begin{conj}
Within a given cochain complex at fixed $h$, any state transforming in the largest ${\rm SU}(2)_a\times {\rm SU}(2)_b$ representation is fortuitous.
\end{conj}

For example, consider the states in $V_{(1^{N-2},2),0}^1$ with $(h,j)=(2,0)$ that transform in the largest possible representation, $({\bf 5},{\bf 4})$. We find that these states carry excitations only on the length-2 cycle, in agreement with the argument above. Under the action of $Q$, such states can map into the space $V_{(1^N),0}^2\oplus V_{(1^{N-3},3),0}^2$. However, the subspace $V_{(1^N),0}^2$ does not contain any state transforming in the representation $({\bf 1},{\bf 2})\otimes({\bf 5},{\bf 4})=({\bf 5},{\bf 3})\oplus({\bf 5},{\bf 5})$. Hence, the states in the $({\bf 5},{\bf 4})$ representation are $Q$-closed at $N=2$ and are therefore fortuitous.

\subsubsection*{R-charge concentration} 

The R-charge concentration conjecture of \cite{Chang:2024lxt} proposes that in each irreducible cochain complex, all fortuitous states concentrate in a single space, while the monotone states may spread among all the spaces in the cochain complex. Our previous results on the $N=2$ fortuitous classes clearly exhibit the R-charge concentration.

We further argue that the $Q_3$-cohomology computed by the longest cochain complex also obey the R-charge concentration. In non-central spaces $V_{(1^3),0}^{0}$, $V_{(1,2),0}^{1}$, $V_{(1,2),0}^{3}$, $V_{(1^3),0}^{4}$ in the $N=3$ complex in \eqref{eqn:longestcochain}, by explicit computations up to $h=4$, we find that all the $\frac14$-BPS states are contracted large-$\mathcal{N}=4$ descendants of half-BPS states;\footnote{The degeneracy of BPS states in $V_{(1^3),0}^{0},V_{(1^3),0}^{4}$ is captured by $\chi_0$, while the degeneracy in $V_{(1,2),0}^{1}$, $V_{(1,2),0}^{3}$ is captured by $\chi_1+2\chi_2$. The explicit form of $\chi_0,\chi_1,\chi_2$ is provided in appendix \ref{subsec: N=3characters}.} hence, are all monotone. In other words, the fortuitous classes up to $h=4$ must reside in the central spaces $V^3_{(3),0}$ and $V^3_{(1^3),0}$.

\section{Outlook}

The ${\rm SU}(2)_a \times {\rm SU}(2)_b$ symmetry plays a central role in our analysis of fortuitous states: within a given cochain complex, we conjecture that states transforming in the largest representation are fortuitous. This conjecture is reminiscent of recent observations in the study of fortuity in the single-matrix model~\cite{Chen:2025sum}. Moreover, the ${\rm SU}(2)_a \times {\rm SU}(2)_b$ symmetry can be used to further refine the modified index and to probe the distribution of representations at larger $N$ using index-based methods~\cite{Benjamin:xxx}.

Our results provide strong evidence for the R-charge concentration phenomenon at $N=2$ and 3: within each cochain complex, the BPS states are supported in the central space, which also has the maximal dimension in the complex. At large but finite $N$, the supercharge cochain complex becomes highly intricate and involves states with complicated cycle shapes, and from the CFT perspective it remains unclear whether this concentration pattern persists in the strict large-$N$ limit. If the observed correlation between R-charge concentration and the dimensions of the constituent spaces continues to hold, a sharper understanding of the dimension growth along the complex becomes a natural next step.

Our analysis in this paper focuses on the longest supercharge cochain complex; systematic studies of the remaining complexes will appear in~\cite{Chang:xxx}. The central space in the longest complex contains all maximally twisted states, which are expected to be dual to single-center black-hole microstates~\cite{Maldacena:1999bp}. There is a large class of other bulk solutions, such as black rings, multi-center bubbled geometries, and multi-center black holes~\cite{Emparan:2001wn,Gauntlett:2002nw, Elvang:2004ds,Elvang:2004rt, Bena:2005ay, Bena:2007kg,Candlish:2009vy, Gibbons:2013tqa,Lucietti:2020ryg,Bena:2022rna,Bena:2025pcy}, whose precise boundary duals remain to be identified. A natural possibility is that they are dual to non-maximally twisted fortuitous states residing in the central spaces of the other cochain complexes.

\section*{Acknowledgements} 
We would like to thank Iosif Bena, Nathan Benjamin, Yiming Chen, Matthias Gaberdiel, Bin Guo,  Marcel Hughes, Ho tat Lam, Ji Hoon Lee, Wei Li,  Masaki Shigemori, Nick Warner, and Zhenbin Yang for helpful discussions, and Ying-Hsuan Lin for collaboration in other projects. CC is partly supported by the National Key R\&D Program of China (NO. 2020YFA0713000). HZ thanks the hospitality of Yau Mathematical Sciences Center,
where part of the work was done during the visit. This work was performed in part at the Aspen Center for Physics, which is supported by National Science Foundation grant PHY-2210452.

\appendix
\section{Orbifold character}
\subsection{Short characters at $N=2$}
Following \cite{Eguchi:1987wf, Eguchi:1987sm}, there are two short characters at $N=2$, which we denote by 
\be
\chi_0, \chi_1
\ee
Their expansion takes the following form
\ie\label{eqn:N=2_short_char_j=0}
\chi_{0}=&\frac{1}{q^{\frac12}}-2\left(y+\frac{1}{y}\right)+q^{\frac12} \left(2 y^2+\frac{2}{y^2}+9\right)-2 q\left(y^3+9 y+\frac{9}{y}+\frac{1}{y^3}\right)
\\
&+q^{\frac32}
   \left(y^4+22 y^2+60+\frac{22}{y^2}+\frac{1}{y^4}\right)-2 q^2 \left(9 y^3+55 y+\frac{55}{y}+\frac{9}{y^3}\right)
\\
&+q^{\frac52} \left(9 y^4+132
   y^2+305+\frac{132}{y^2}+\frac{9}{y^4}\right)
-2 q^3 \left(y^5+55 y^3+266 y+\frac{266}{y}+\frac{55}{y^3}+\frac{1}{y^5}\right)\\
& + q^{\frac{7}{2}}(60y^4+60 \frac{1}{y^4}+634 y^2+634y^{-2}+1323)+O\left(q^{4}\right)\,,
\fe
and
\ie\label{eqn:N=2_short_char_j=1}
\chi_1=&-\left(y+\frac1y\right)+2 q^{\frac12} \left(y^2+3+\frac{1}{y^2}\right)-q\left(y^3+15 y+\frac{15}{y}+\frac{1}{y^3}\right)
\\
&+q^{\frac32} \left(20 y^2+54+\frac{20}{y^2}\right)-q^2 \left(15 y^3+113 y+\frac{113}{y}+\frac{15}{y^3}\right)
\\
&+q^{\frac 52} \left(6 y^4+144 y^2+338+\frac{144}{y^2}+\frac{6}{y^4}\right)
-q^3 \left(y^5+113 y^3+630 y+\frac{630}{y}+\frac{113}{y^3}+\frac{1}{y^5}\right)\\
&+q^{7/2}(54y^4+54y^{-4}+772y^2+772y^{-2}+1662)+O\left(q^{4}\right)\,.
\fe
\subsection{Characters at $N=3$}\label{subsec: N=3characters}
At $N=3$, there are three short characters, which we denote by  
\[
\chi_{0}, \quad \chi_{1}, \quad \chi_{2}. \label{eqn: N=3 short characters}
\]  
Their expansion takes the form 
\bea
&&\chi_{0}=\frac{1}{q^{\frac{3}{4}}}\big[-1 + 2 q^{\frac{1}{2}} (y+\frac{1}{y}) - q (9+2 y^{-2}+2y^2) + q^{\frac{3}{2}}(2y^{-3}+18y^{-1}+18y+2y^3) \crcr
&&- q^2 (2y^{-4}+23y^{-2}+61+23 y^2 +2y^4)+ q^{\frac{5}{2}}(2 y^{-5}+24 y^{-3}+116 y^{-1}+ 116 y +24 y^3 +2 y^5)\crcr
&&-q^3(y^{-6}+23y^{-4}+153y^{-2}+326+153y^2+23y^4+y^6)\crcr
&&+q^{\frac{7}{2}}(18y^{-5}+164y^{-3}+594y^{-1}+594y+164y^3+18y^5)\crcr
&&-q^4(9y^{-6}+153y^{-4}+795y^{-2}+1492+795y^2+153y^4+9y^6)+O(q^{\frac{9}{2}}) \big]
\eea
\bea
&&\chi_{1}=\frac{1}{q^{\frac{3}{4}}}\big[ q^{\frac{1}{2}} (y+\frac{1}{y}) - q (6+2 y^{-2}+2y^2) + q^{\frac{3}{2}}(2y^{-3}+16y^{-1}+16y+2y^3) \crcr
&&- q^2 (2y^{-4}+26y^{-2}+60+26 y^2 +2y^4)+ q^{\frac{5}{2}}(y^{-5}+30 y^{-3}+135 y^{-1}+ 135 y +30 y^3 + y^5)\crcr
&&-q^3(26y^{-4}+206y^{-2}+408+206y^2+26y^4)+q^{\frac{7}{2}}(16y^{-5}+234y^{-3}+826y^{-1}+826y +234y^3+16y^5)\crcr
&&-q^4(6y^{-6}+206y^{-4}+1214y^{-2}+2192+1214y^2+206y^4+)+O(q^{\frac{9}{2}}) \big]
\eea
\bea
&&\chi_{2}=\frac{1}{q^{\frac{3}{4}}}\big[ - q (1+ y^{-2}+y^2) + q^{\frac{3}{2}}(2y^{-3}+6y^{-1}+6y+2y^3) \crcr
&&- q^2 (y^{-4}+15^{-2}+22+15 y^2 +1y^4)+ q^{\frac{5}{2}}(20 y^{-3}+62 y^{-1}+ 62 y +20 y^3)\crcr
&&-q^3(15y^{-4}+121y^{-2}+188+121  y^2+15 y^4)+q^{\frac{7}{2}}(6 y^5 +152y^3+434y+434y^{-1}+152y^{-3}+6y^5) \crcr
&&-q^4(y^{-6}+121y^{-4}+733y^{-2}+1144+733y^2+121y^4+y^6)+O(q^{\frac{9}{2}})\big]
\eea

In addition, there are long characters of two distinct structures, given by  
\[
\chi_{h,0} \;=\; q^h \bigl(\chi_{0} + 2\chi_{1} + \chi_{2}\bigr), 
\qquad 
\chi_{h+\tfrac{1}{2},1} \;=\; q^h \bigl(\chi_{1} + 2\chi_{2}\bigr).
\]

\section{Cochain Complexes in the $N=2$ and $3$ theories}\label{appendix: cochain_complex_def}
In this Appendix, we spell out the cochain complexes computing the $Q$-cohomology in the $N=2$ and $N=3$ theories and explain how they are related via the projection map $\pi_{2,3}$. The cochains in these complexes are spaces of $\tfrac14$-BPS states at the orbifold point, and the differential is the first-order deformed supercharge $Q$, which is invariant under the right-moving ${\rm SU}(2)_{\rm outer}$ symmetry and commutes with the right-moving diagonal Clifford algebra \cite{Chang:2025rqy}. A Clifford quartet takes the form
\ie
\ket{\Omega}\,,\quad \tilde\psi^{+\rm diag}_{-\frac12}\ket{\Omega}\,,\quad \tilde{\bar\psi}^{+\rm diag}_{-\frac12}\ket{\Omega}\,,\quad \tilde\psi^{+\rm diag}_{-\frac12}\tilde{\bar\psi}^{+\rm diag}_{-\frac12}\ket{\Omega}\,,
\fe
where $\tilde\psi^{+\rm diag}_{-\frac12}$ and $\tilde{\bar\psi}^{+\rm diag}_{-\frac12}$ are the raising operators of the right-moving diagonal Clifford algebra, and $\ket{\Omega}$ is the bottom component of the Clifford quartet, defined as the state annihilated by the lowering operators. Suppose $\ket{\Omega}$ transforms in the representation ${\bf n}$ of ${\rm SU}(2)_{\rm outer}$; then the Clifford quartet transforms in the tensor product representation $({\bf 1}\oplus{\bf 2}\oplus{\bf 1}) \otimes {\bf n}$. In what follows, we focus on the bottom components of the Clifford quartets and project onto the highest-weight states under ${\rm SU}(2)_{\rm outer}$. The resulting cochain complexes are
\ie
\begin{tikzcd}
 0 \arrow[r, "Q_{3}"] 
   & V_{(1^3),0}^{0} \arrow[r, "Q_{3}"] \arrow[d, "\pi_{2,3}"]  
   & V_{(1,2),0}^{1}\arrow[r, "Q_{3}"] \arrow[d, "\pi_{2,3}"]  
   & V_{(1^3),0}^{2} \oplus
 V_{(3),0}^{2} \arrow[r, "Q_{3}"] \arrow[d, "\pi_{2,3}"] 
    & V_{(1,2),0}^{3}\arrow[r, "Q_{3}"] \arrow[d, "\pi_{2,3}"] & V_{(1^3),0}^{4}\arrow[d, "\pi_{2,3}"]  \arrow[r, "Q_{3}"]  \arrow[d, "\pi_{2,3}"] & 0 \\
 0 \arrow[r, "Q_2"] 
   & V_{(1^2),0}^{0} \arrow[r, "Q_2"]  
   & V_{(2),0}^{1} \arrow[r, "Q_2"]  
   & V_{(1^2),0}^{2}  \arrow[r, "Q_2"]   & 0 & 0 & 
\end{tikzcd}
\fe
\ie
\begin{tikzcd}
 0 \arrow[r, "Q_{3}"] 
   & V_{(1^3), 1}^{1} \arrow[r, "Q_{3}"] \arrow[d, "\pi_{2,3}"]  
   & V_{(1,2),1}^{2}\arrow[r, "Q_{3}"]  \arrow[d, "\pi_{2,3}"]  
   & V_{(1^3),1}^{3} \arrow[r, "Q_{3}"] \arrow[d, "\pi_{2,3}"]  & 0 \\
 0 \arrow[r, "Q_{2}"]  
   & V_{(1,1),1}^1 \arrow[r, "Q_2"]  
   & 0 
   & 0      & 
\end{tikzcd}
\fe
\ie
\begin{tikzcd}
 0 \arrow[r, "Q_{3}"] 
   & V_{(1^3),2}^{2} \arrow[r, "Q_{3}"] \arrow[d, "\pi_{2,3}"]  
   & 0 \\
   & 0  
   & 
\end{tikzcd}
\fe
where the cochains $V^{2\tilde{\jmath}}_{p,\,2\tilde{\jmath}_{\text{outer}}}$ are labeled by the cycle shape $p$, right-moving $R$-charge $\tilde{\jmath}$ and right-moving $\mathrm{SU}(2)_{\text{outer}}$ spin $\tilde{\jmath}_{\text{outer}}$.

The bottom of $V_{(1^3),0}^{0}, V_{(1^3),0}^{2},V_{(1^3),0}^{4}$ are spanned by 
\bea
&& \sum_{g\in S_3} g  \, \ket{\Psi_{123}}\crcr
&&\sum_{g\in S_3} g \, (\tilde\psi^{[1]+}_{-\frac12}-\tilde\psi^{[2]+}_{-\frac12})(\tilde{\bar\psi}^{[1]+}_{-\frac 12}-\tilde{\bar\psi}^{[2]+}_{-\frac 12})\ket{\Psi_{(12)3}}\crcr
&&    \sum_{g\in S_3} g \, ((\bar\psi^{[1]+}_{-\frac{1}{2}}\psi^{[2]+}_{-\frac{1}{2}}-\psi^{[1]+}_{-\frac{1}{2}}\bar\psi^{[2]+}_{-\frac{1}{2}}+\frac{1}{3}[\psi^{[2]+}_{-\frac{1}{2}}\bar\psi^{[2]+}_{-\frac{1}{2}}+\psi^{[1]+}_{-\frac{1}{2}}\bar\psi^{[1]+}_{-\frac{1}{2}}])\psi^{[3]+}_{-\frac{1}{2}}\bar\psi^{[3]+}_{-\frac{1}{2}})\ket{\Psi_{123}}  
\eea
where $\ket{\Psi_{123}}$ denotes possible excitations of the tensor product of three $T^4$ theory. 

The bottom of $V_{(1,2),0}^1$ and $V_{(1,2),0}^3$ are spanned by 
\bea
&&\sum_{g\in S_3} \ket{\Psi_{123}}\crcr
&&\sum_{g\in S_3}  (\sqrt{\frac{2}{3}}\tilde{\psi}^{[1]+}_{-\frac{1}{2}}-\sqrt{\frac{1}{3}}\tilde{\psi}^{[23]+}_{-\frac{1}{2}})(\sqrt{\frac{2}{3}}\tilde{\psi}^{[1]+}_{-\frac{1}{2}}-\sqrt{\frac{1}{3}}\tilde{\psi}^{[23]+}_{-\frac{1}{2}})\ket{\Psi_{123}}
\eea

$V_{(1^3),1}^{1}$ contains two right-moving quartets, whose bottom components are spanned by
\bea
&&\sum_{g\in S_3}(\tilde{\psi}^{[1]+}_{-\frac{1}{2}} -\tilde{\psi}^{[2]+}_{-\frac{1}{2}})\ket{\Psi_{123}}\crcr
&&\sum_{g\in S_3}(\tilde{\bar\psi}^{[1]+}_{-\frac{1}{2}} -\tilde{\bar\psi}^{[2]+}_{-\frac{1}{2}})\ket{\Psi_{123}}
\eea
$V_{(1^3),1}^3$ contains two right-moving quartets, whose bottom components are spanned by
\bea
&&\sum_{g\in S_3}(\tilde{\psi}^{[1]+}_{-\frac{1}{2}} -\tilde{\psi}^{[2]+}_{-\frac{1}{2}})(\tilde{\bar\psi}^{[1]+}_{-\frac{1}{2}} -\tilde{\bar\psi}^{[2]+}_{-\frac{1}{2}})(\frac{1}{2}\tilde{\psi}^{[1]+}_{-\frac{1}{2}} +\frac{1}{2}\tilde{\psi}^{[2]+}_{-\frac{1}{2}}- \tilde{\psi}^{[3]+}_{-\frac{1}{2}})\ket{\Psi_{123}}\crcr
&&\sum_{g\in S_3}(\tilde{\psi}^{[1]+}_{-\frac{1}{2}} -\tilde{\psi}^{[2]+}_{-\frac{1}{2}})(\tilde{\bar\psi}^{[1]+}_{-\frac{1}{2}} -\tilde{\bar\psi}^{[2]+}_{-\frac{1}{2}})(\frac{1}{2}\tilde{\bar\psi}^{[1]+}_{-\frac{1}{2}} +\frac{1}{2}\tilde{\bar\psi}^{[2]+}_{-\frac{1}{2}}- \tilde{\bar\psi}^{[3]+}_{-\frac{1}{2}})\ket{\Psi_{123}}
\eea
$V_{(1,2),1}^{2}$ contain two quartets, whose bottom components are spanned by
\bea
&&\sum_{g\in S_3}g(\sqrt{\frac{1}{3}}\tilde{\psi}^{[1]+}_{-\frac{1}{2}} - \sqrt{\frac{2}{3}}\tilde{\psi}^{[23]+}_{-\frac{1}{2}})\ket{\Psi_{123}}\crcr
&&\sum_{g\in S_3}g(\sqrt{\frac{1}{3}}\tilde{\bar\psi}^{[1]+}_{-\frac{1}{2}} - \sqrt{\frac{2}{3}}\tilde{\bar\psi}^{[23]+}_{-\frac{1}{2}})\ket{\Psi_{123}}
\eea
$V_{(1^3),2}^{2}$ contains three totally anti-symmetric right-moving quartets, whose bottom components are spanned by
\bea
&&\sum_{g\in S_3}(\tilde{\psi}^{[1]+}_{-\frac{1}{2}} -\tilde{\psi}^{[2]+}_{-\frac{1}{2}})\tilde{\bar\psi}^{[3]+}_{-\frac{1}{2}}+(\tilde{\psi}^{[2]+}_{-\frac{1}{2}} -\tilde{\psi}^{[3]+}_{-\frac{1}{2}})\tilde{\bar\psi}^{[1]+}_{-\frac{1}{2}}+(\tilde{\psi}^{[3]+}_{-\frac{1}{2}} -\tilde{\psi}^{[1]+}_{-\frac{1}{2}})\tilde{\bar\psi}^{[2]+}_{-\frac{1}{2}}\ket{\Psi_{123}}\crcr
&&\sum_{g\in S_3}(\tilde{\psi}^{[1]+}_{-\frac{1}{2}} \tilde{\psi}^{[2]+}_{-\frac{1}{2}}-\tilde{\psi}^{[1]+}_{-\frac{1}{2}} \tilde{\psi}^{[3]+}_{-\frac{1}{2}}+\tilde{\psi}^{[2]+}_{-\frac{1}{2}} \tilde{\psi}^{[3]+}_{-\frac{1}{2}}) \ket{\Psi_{123}}\crcr
&&\sum_{g\in S_3}(\tilde{\bar\psi}^{[1]+}_{-\frac{1}{2}} \tilde{\bar\psi}^{[2]+}_{-\frac{1}{2}}-\tilde{\bar\psi}^{[1]+}_{-\frac{1}{2}} \tilde{\bar\psi}^{[3]+}_{-\frac{1}{2}}+\tilde{\bar\psi}^{[2]+}_{-\frac{1}{2}} \tilde{\bar\psi}^{[3]+}_{-\frac{1}{2}})\ket{\Psi_{123}}
\eea

\section{${\rm SU}(2)_a\times {\rm SU}(2)_b$ charges}
\label{app:SU2axSU2b_symmetry}

As discussed in \cite{Gaberdiel:2025smz}, the deformation preserves two additional charges in addition to the SU(2) 
R-charge. We further note that these two charges can be enhanced to two 
SU(2) zero mode charge. The explicit modes for the left-moving sector of zero mode are presented in the following
\bea
&&J_1^b =\frac{1}{2 i} \sum_{n} \frac{1}{n}(\bar\alpha^1_{-n} \alpha^1_n+ \bar\alpha^2_{-n}\alpha^2_n )\crcr
&&J_2^b =\frac{1}{2 } \sum_{n} \frac{1}{n}(\bar\alpha^2_{-n}\alpha^2_n- \bar\alpha^1_{-n} \alpha^1_n )\crcr
&& J_3^b= \frac{1}{2} \sum_{n} \frac{1}{n}(
\bar\alpha^2_{-n} \alpha^1_n+ \bar\alpha^1_{-n}\alpha^2_n )
\eea

\bea
&&J_1^a =\frac{1}{2 i}[ \sum_{n} \frac{1}{n}(\bar\alpha^2_{-n} \bar\alpha^1_n+ \alpha^2_{-n}\alpha^1_n) - \psi^-_{-n}\psi^+_{n} + \bar\psi^-_{-n} \bar\psi^+_{n} ]\crcr
&&J_2^a =\frac{1}{2 } \sum_{n}[ \frac{1}{n}(\bar\alpha^2_{-n} \bar\alpha^1_n- \alpha^2_{-n}\alpha^1_n )+ \psi^-_{-n}\psi^+_{n} + \bar\psi^-_{-n} \bar\psi^+_{n} ]\crcr
&& J_3^a= \frac{1}{2} \sum_{n} [\frac{1}{n}(
\bar\alpha^2_{-n} \alpha^1_n+ \alpha^2_n\bar\alpha^1_{-n})-\bar\psi^+_{-n} \psi^-_n + \psi^+_{-n}\bar\psi^-_n ]
\eea

\section{Representation decomposition of $h=3,j=0$ BPS states}\label{app:RepDecompofh=3j=0}
We find the $SU(2)_a\times SU(2)_b$ representation decomposition of $\frac{1}{4}$-BPS states at $h=3,j=0$ take the following form. 
\begin{table}[H]
\begin{center}
\begin{tabular}{|c|c|c|}
\hline
Representation & Degeneracy \\ \hline\hline
$ ({\bf 6},{\bf 5}) $ & $\textcolor{red}{3}$   \\
$({\bf 4},{\bf 5})$ & $\textcolor{red}{3}$  \\
$  ({\bf 6},{\bf 3})$ & $\textcolor{red}{3}$  \\
$  ({\bf 4},{\bf 3})$ & $9= \textcolor{red}{6}+3$  \\
$ ({\bf 2},{\bf 3})$ & $13=\textcolor{red}{10}+3$  \\
$ ({\bf 4},{\bf 1})$ & $6=\textcolor{red}{3}+3 $ \\
$  ({\bf 2},{\bf 1})$ & $37=\textcolor{red}{3}+34$  \\
\hline
\end{tabular}~~~~~~~~~~
\begin{tabular}{|c|c|c|}
\hline
Representation & Degeneracy \\ \hline\hline
$ ({\bf 7},{\bf 6})$ & $\textcolor{red}{1}$   \\
$  ({\bf 5},{\bf 6})$ & $\textcolor{red}{1}$  \\
$ ({\bf 7},{\bf 4})$ & $\textcolor{red}{1}$  \\
$ ({\bf 5},{\bf 4})$ & $ \textcolor{red}{5}$  \\
$ ({\bf 3},{\bf 4})$ & $ 4=\textcolor{red}{3}+1$  \\
$ ({\bf 5},{\bf 2})$ & $\textcolor{red}{2}$  \\
$ ({\bf 3},{\bf 2})$ & $22=\textcolor{red}{6}+16 $ \\
$ ({\bf 3},{\bf 2})$ & $29=\textcolor{red}{2}+27$  \\
\hline
\end{tabular}
\end{center}
\caption{}
\end{table}
We use red color to denote the degeneracy of fortuitous states, and the black color denotes the degeneracy of monotone states. Note that the fortuitous states and monotone states presented here are a combination of left-moving contracted $\mathcal{N}=(4,4)$ primaries and descendants. 

\bibliography{refs}

\bibliographystyle{utphys}

\end{document}